# Status of the SPHERE experiment

R A Antonov[1], S P Beschapov[2], E A Bonvech[1], D V Chernov[1], T A Dzhatdoev[1], Mir Finger[3], M Finger[3], V I Galkin[1,4], N N Kabanova[2], A S Petkun[2], D A Podgrudkov[1,4], T M Roganova[1], S B Shaulov[2] and T I Sysoeva[2]

[1] Lomonosov Moscow State University Skobeltsyn Institute of Nuclear Physics, Russia
[2] P.N. Lebedev Physical Institute of the Russian Academy of Sciences, Russia
[3] Charles University in Prague, Czech Republic
[4] Physics Department of Lomonosov Moscow State University, Russia
E-mail: chr@dec1.sinp.msu.ru

**Abstract**. Here is presented the current state of the SPHERE-2 balloon-borne experiment. The detector is elevated up to 1 km above the snow surface and registers the reflected Vavilov-Cherenkov radiation from extensive air showers. This method has good sensitivity to the mass-composition of the primary cosmic rays due to its high resolution near the shower axis. The detector consists of a 1500 mm spherical mirror with a 109 PMT cluster in its focus. The electronics record a signal pulse profile in each PMT. In the last 2 years the detector was upgraded: time resolution of pulse registration was enhanced up to 12.5 ns, channel sensitivity was increased by a factor of 3, a new LED-based relative PMT calibration method was introduced, and new hardware and etc. was installed.

## 1. Introduction
The SPHERE experiment is based on the method first proposed by A.E. Chudakov [1]. The Vavilov-Cherenkov radiation generated by the extensive air shower (EAS) is reflected by the snow surface and registered by the detector lifted by the tied balloon to the altitude ranging from several hundred meters up to several kilometers [2]. The detector system works like a camera and registers the images of Vavilov-Cherenkov radiation spots produced by EAS. At the altitudes up to 1 km the detector can register the EAS images corresponding to primary cosmic rays with energies about 10 – 500 PeV.

## 2. The SPHERE-2 detector
The SPHERE-2 detector consist of a 1500 mm diameter and 940 mm curvature radius seven segment spherical mirror and the mosaic of 109 PMTs FEU-84-3 (figure 1) on the mirror focal surface. PMTs have the hexagonal arrangement in the mosaic (figure 2). Diameter of FEU-83-3 photocathode sensitive area is 28 mm. To enhance the optical resolution and keep the detectors field of view and aperture the diaphragm (diameter 940 mm) is placed in front of the mirror. The detector field of view is 1 sr.

Electronic part of the detector consists of data acquisition system (DAQ), trigger system (TS), calibration system (CS) and control block. DAQ contains 109 pulse shape acquisition 10-bit FADC channels with 12.5 ns steps. Trigger system allows detection on the PMTs mosaic of a Vavilov-Cherenkov radiation spot image (at least 3 adjacent PMTs have signal above threshold within 1 μs interval). CS contains 7 UV LEDs that produce independently controlled light pulses over the mosaic.

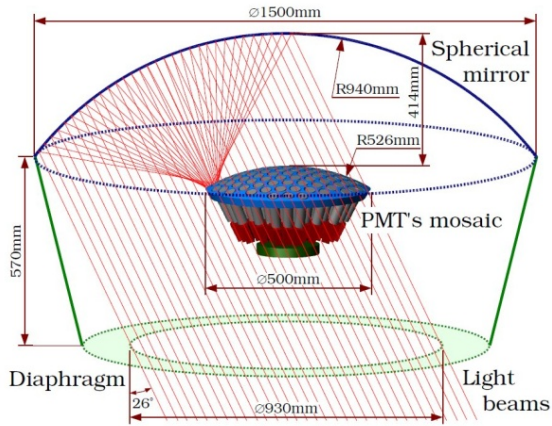 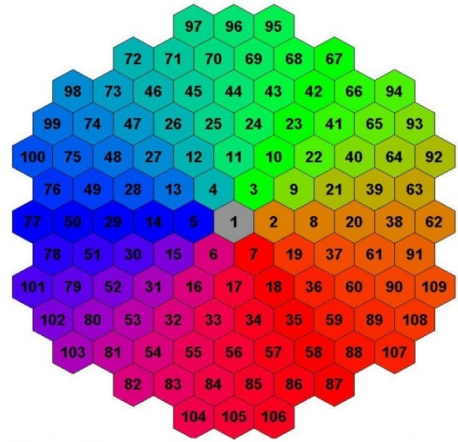

Figure 1. Optical scheme of SPHERE-2 detector.   Figure 2. The location of PMTs in the mosaic.

For the preliminary analysis of the registered events their graphic representation similar to the one given in the figure 3 is used. Each pixel corresponds to one measurement of the signal in one PMT with a step of 12.5 ns. Color indicates the measured intensity. The picture shows that a physical event forms a sine-like curve in such oscillogram representation. This is due to the numeration of the channels in the PMT' mosaic. Each subsequent period of the curve in the figure corresponds to signals from PMTs in the same "ring" on the surface of the mosaic (the first "ring" is formed by PMTs NN 2–7, the second "ring" – NN 8–19 and so on). The time amplitude of the curve correlates with the zenith angle of the EAS axis. Phase shift of the curve reflects the azimuthal angle of the EAS axis in detector coordinates (the angle between the axis projection and a line connecting PMT 1 and PMT 62). The high intensity pulses (those with a signal to noise ratio above 5 – triangles on figure 3a) are used to detect the shower plane and give an estimation for the low intensity pulse search (squares on figure 3b). Each pulse after being found is fitted with a pulse function. This allows to work with very weak pulses (S/N ~ 1 or even less).

The given method of analysis is applied for preliminary estimations of the EAS parameters, and also as a characteristic of the overall performance of equipment. Estimations of the EAS parameters ($\Theta$, $\varphi$, $X_0$, $Y_0$, $E$) are made by minimization of several functionalities [3]. The direction of an EAS axis is calculated with the assumption of the flat front of Vavilov-Cherenkov radiation of EAS (figure 4).

*a* *b*

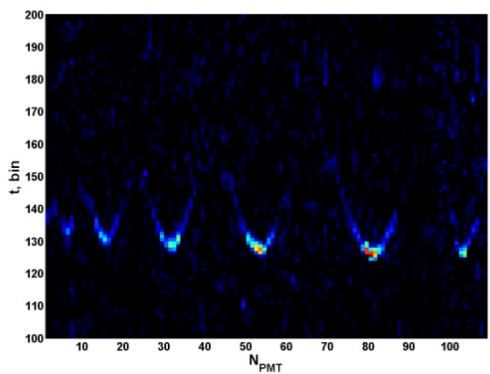 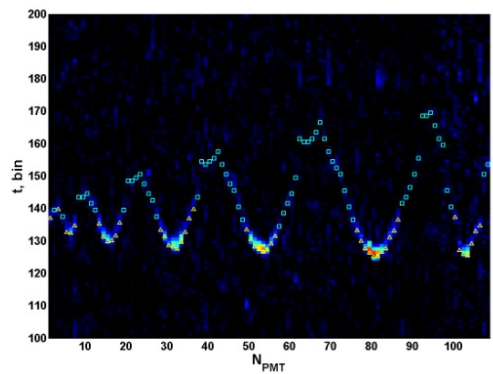

Figure 3. Detected event representation. $N_{PMT}$ - number of PMT in mosaic (see figure 2), t – relative time of PMT's signal (1 bin = 12.5 ns) *a)* input event signals, *b)* the same signals, triangles – high signal labels used for weak signal positions reconstruction (squares).

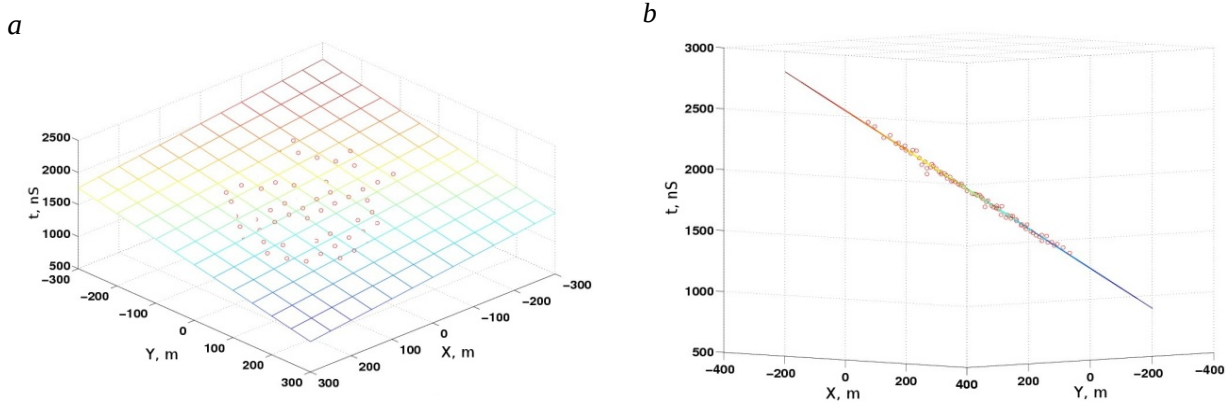

Figure 4. Angular reconstruction of experimental event by a signal time delay analysis. Circles represent time delay of a signal maximum in different PMTs from the corresponding areas on a snow surface. On *a* and *b* the same event is shown from different angles.

On the figure 5 examples of Cherenkov LDFs are presented. They show that even at relatively small elevations the detector shows high spatial resolution. The nearest to the axis point is only 25 m away. The field of view of the PMT corresponding to this point observes an area that contains the EAS axis. Measurements of the density of Vavilov-Cherenkov photons near the EAS axis is practically impossible by the majority of ground-based arrays. However the near-axis region is the most sensitive part of the EAS to the primary particle type.

The event analysis includes the comparison of the experimental data with detector responses (optics included) from modeled showers. However, for the preliminary event reconstruction an LDF fitting procedure is used. The approximation from paper [4] was used (with author's comments and corrections given in private communications). It's a two parameter piecewise function with a detailed near-axis (less than 200 m) structure. Originally this approximation fits the exact Vavilov-Cherenkov photons LDF (without any optical effects), but it shows good correlation with our data and thus is appropriate for preliminary analysis. Figures show the shower's complex structure near the axis. The features of this structure are sensitive to the primary particle type.

An energy estimation can be obtained using some traditional parameters (photon densities at a certain distance from the shower axis) or, as was done in this analysis, using the full number of Vavilov-Cherenkov photons for which the approximation [4] helps greatly.

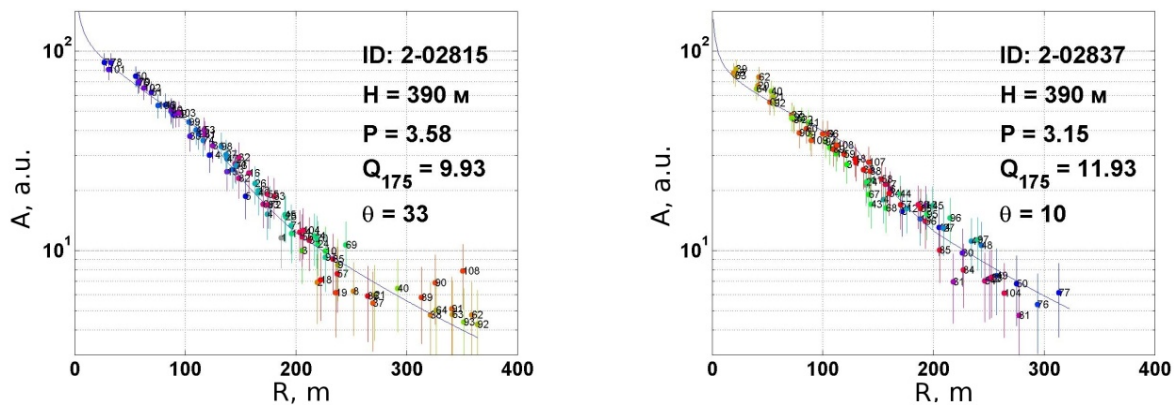

Figure 5. LDF of two events from experimental data of March 2012. Colour dots and numbers corresponds to the PMT position in mosaic on figure 2, the curves are LDF approximation (see text)

## 3. Features of measurements in 2012

Before the expedition of 2012 the electronics of the detector have been essentially improved. The stability of DAQ was raised due to a new program for the FPGA chips. The sensitivity of the preamplifiers was increased by 3 times. Two ADC on each measuring channel, which earlier worked synchronously but with different amplification factors, now have identical amplification factors and a half-phase shift. This allows to double the time resolution to 12.5 ns. Thus the accuracy of the anode pulse charge digitizing was increased by 6 times. A calibration system was added, which enhanced the precision of all time-related measurements by 3-4 times and gave the PMTs relative calibration with an accuracy of about 3%.

An electronic compass for measuring the azimuth of the detector position was mounted. Together with the GPS receiver and the inclinometer mounted earlier, this system allows to determine the detector position relative to the Earth and to reconstruct the direction of the PCR in astronomical coordinates.

## 4. Conclusion

There is significant progress in techniques of measurement and reconstruction of the EAS parameters from the SPHERE-2 experimental data. As a result it has became clear that the detector has some advantages in respect to the ground-based Cherenkov arrays. First, each of the SPHERE-2 detectors gathers light from a large area what makes the experiment insensitive to local photon density fluctuations. Second, it is capable of measuring the density of Vavilov-Cherenkov photons near the axis of almost every EAS thanks to the quasi-continuous sensitive surface of the detector. These two points allow to significantly improve the reconstruction accuracy of the lateral distribution function of the Vavilov-Cherenkov radiation and to increase the sensitivity of the detector to the chemical composition of the primary cosmic rays.

By changing the altitude of SPHERE-2 over the surface of the snow allows to study LDF with high spatial resolution (up to 30 m at a height of about 350 m). The energy range of the measured EAS rises by up to 10 times at an elevation of 1000 m and more. Exposure time is significantly increased.

The small size of the SPHERE-2 detector allows to simplify equipment and to reduce the experimental cost in comparison with the ground-based arrays.

## 5. Acknowledgments

The authors acknowledge the Russian Foundation for Basic Research (grant 11-02-01475-a, 12-02-10015-k) and the Program of basic researches of the Presidium of the Russian Academy of Sciences "Fundamental properties of matter and astrophysics" for the support of the research.